\documentclass[aip,jcp,reprint,showpacs,showkeys,superscriptaddress,longbibliography]{revtex4-1} % nobalancelastpage, endfloats*, linenumbers
\usepackage[utf8]{inputenc}
\usepackage[UKenglish]{babel}

\usepackage{cmap} % make ligatures etc. copy- and searchable
\usepackage[T1]{fontenc}
\usepackage{microtype}
\usepackage{amsmath,mathtools,mleftright}

\IfFileExists{mtpro2.sty}
{ % use MathTime Professional 2
  \usepackage{times}
  \usepackage[mtphrb,mtpscr,subscriptcorrection,slantedGreek,zswash]{mtpro2} % mtpcal
}
{ % % searching the PDF for ligatures like in "diffusion" does not work with the
  \usepackage{txfonts,amssymb}
  
  \usepackage[scr=boondoxo]{mathalfa} % \mathscr used for calligraphic symbol 'N'
}
\usepackage[scr=false]{rsfso} % calligraphic math font Formal Script, likely to be used by J Chem Phys

\usepackage{graphicx,grffile,textcomp}
\graphicspath{{figures/}}
\newlength{\figwidth}

\usepackage{siunitx}

\usepackage{url,hyperref}
\usepackage[usenames,dvipsnames]{xcolor}
\hypersetup{colorlinks=true, linkcolor=BrickRed, urlcolor=blue!50!black, citecolor=blue!50!black}

\usepackage[capitalize]{cleveref}
\crefrangelabelformat{equation}{(#3#1#4)$-$(#5#2#6)}

\renewcommand\epsilon{\varepsilon}
\renewcommand\phi{\varphi}
\renewcommand\theta{\vartheta}
\renewcommand\rho{\varrho}
\renewcommand\leq{\leqslant}
\renewcommand\geq{\geqslant}
\renewcommand\vec[1]{\textrm{\bfseries #1}}
\newcommand\diff{\mathrm{d}}
\newcommand\expect[1]{\mleft\langle{#1}\mright\rangle}
\newcommand\e{\text{e}}
\renewcommand\i{\text{i}}

\newcommand\kB{k_{\text{B}}}
\newcommand\Lbox{L_{\text{box}}}
\DeclareMathOperator\Var{var}

\begin{document}

\figwidth=3.4in % about \linewidth (=246.0pt)

\title{Finite-size corrections for the static structure factor of a liquid slab with open boundaries}

\author{F. Höf{}ling}
% \email{f.hoefling@fu-berlin.de}
\affiliation{Freie Universität Berlin, Fachbereich Mathematik und Informatik,
Arnimallee 6, 14195 Berlin, Germany}
\affiliation{Zuse Institute Berlin, Takustr. 7, 14195 Berlin, Germany}
\email{f.hoefling@fu-berlin.de}

\author{S. Dietrich}
\affiliation{Max-Planck-Institut für Intelligente Systeme, Heisenbergstraße 3,
70569 Stuttgart, Germany}
\affiliation{Institut für Theoretische Physik IV,
Universität Stuttgart, Pfaffenwaldring 57, 70569 Stuttgart, Germany}

\date{\today}

\begin{abstract}
\noindent
The presence of a confining boundary can modify the local structure of a liquid markedly.
In addition, small samples of finite size are known to exhibit systematic deviations of thermodynamic quantities relative to their bulk values.
Here, we consider the static structure factor of a liquid sample in slab geometry with open boundaries at the surfaces, which can be thought of as virtually cutting out the sample from a macroscopically large, homogeneous fluid.
This situation is a relevant limit for the interpretation of grazing-incidence diffraction experiments at liquid interfaces and films.
We derive an exact, closed expression for the slab structure factor, with the bulk structure factor as the only input.
This shows that such free boundary conditions cause significant differences between the two structure factors, in particular at small wavenumbers.
An asymptotic analysis of this result yields the scaling exponent and an accurate, useful approximation of these finite-size corrections.
Furthermore, the open boundaries permit the interpretation of the slab as an open system, supporting particle exchange with a reservoir.
We relate the slab structure factor to the particle number fluctuations and discuss conditions under which the subvolume of the slab represents a grand canonical ensemble with chemical potential $\mu$ and temperature $T$.
Thus, the open slab serves as a test-bed for the small-system thermodynamics in a $\mu T$ reservoir.
We provide a microscopically justified and exact result for the size dependence of the isothermal compressibility.
Our findings are corroborated by simulation data for Lennard-Jones liquids at two representative temperatures.
\end{abstract}

% \pacs{68.03.Cd, 61.20.Ja, 05.10.-a}

% 05.10.-a    Computational methods in statistical physics
% 61.20.Ja    Computer simulation of liquid structure
% 61.25.Em    Molecular liquids
% 61.20.Ne    Structure of simple liquids
% 68.03.Hj    Liquid surface structure: measurements and simulations

\maketitle

\section{Introduction}

The present finite-size issue is related to a simulation study of liquid--vapour interfaces \cite{Capillary:2015}, in which two planar slabs of coexisting fluid phases were brought into contact and then into thermal equilibrium.
The goal of that study was to analyse interfacial density fluctuations which are accessible to grazing incidence X-ray diffraction (GIXRD) measurements.
Similar issues arise in studies of liquid--liquid interfaces and liquid films adsorbed on a solid substrate \cite{Rauscher:ARMR2008,Partay:PCCP2008,Gu:FPE2010,Rozas:2011,MacDowell:2017}.
In grazing incidence geometry, the incoming beam hits the liquid surface below the critical angle of total reflection so that merely an exponentially damped, evanescent wave penetrates the liquid bulk phase.
Nevertheless, the scattered intensity picks up a considerable background signal from fluctuations of the bulk phase, which needs to be subtracted in order to expose interface-related fluctuations with non-zero wavenumbers \cite{Scoppola:COCIS2018,Dietrich:1995}.
% scattering review \cite{Scoppola:COCIS2018}
In simulation studies of this kind, one is naturally confined to finite systems so that the penetration depth into the bulk phase is delimited by the system size, which facilitates to replace the exponential damping of the evanescent wave by a sharp cutoff, i.e., a step function.
In either case, the calculation of the background fluctuations rests on the analysis of a macroscopic half-space (for the experiment) or a slab of finite width (for the simulations), featuring an open boundary at the liquid--vapour interface.
Within theoretical treatments such a situation is realised by free boundary conditions for continuous fields describing the physical observables.

Outside this specific context, in recent years,
molecular dynamics (MD) simulations of open systems have received growing interest \cite{Muscatello:2017,Font:JPCC2018,Han:JPCL2017,Bonella:L2017}, % heat transport
one challenge being the study of non-equilibrium phenomena with steady mass transport \cite{Lotfi:2014,Heinen:2016,Wilhelmsen:PRL2015}.
A recently developed methodology, which targets such situations, permits the simulation of a small region of interest coupled via open boundaries to a reservoir, thereby realising an efficient grand canonical sampling of this region while preserving the dynamics \cite{DelleSite:2019}.
Much more directly, open boundaries occur in simulations whenever subsystems of a large simulation domain are considered.
Such open subsystems represent a grand canonical ensemble if the reservoir provided by the remaining domain is sufficiently large \cite{DelleSite:JSM2017,Schnell:CPL2011,Cortes-Huerto:JCP2016,Heidari:E2018}, with a correspondingly high computational cost for the reservoir.
In particular, the statistics of the particle number in the subsystem and the local fluid structure have been shown to be sensitive to the ratio of the reservoir and the subvolume sizes.
In the context of phase transitions, the analysis of subsystems was exploited already much earlier \cite{Binder:1981,Rovere:1993} and taken up recently \cite{Siebert:2018,Chakraborty:2020} in order to determine the properties of Ising-type critical points via finite-size scaling. In such studies, one needs to explicitly account for the open boundary conditions, e.g., by adopting a different value (relative to periodic boundaries) of the critical Binder cumulant.

\begin{figure}[b]
  \centering
  \includegraphics[width=.8\linewidth]{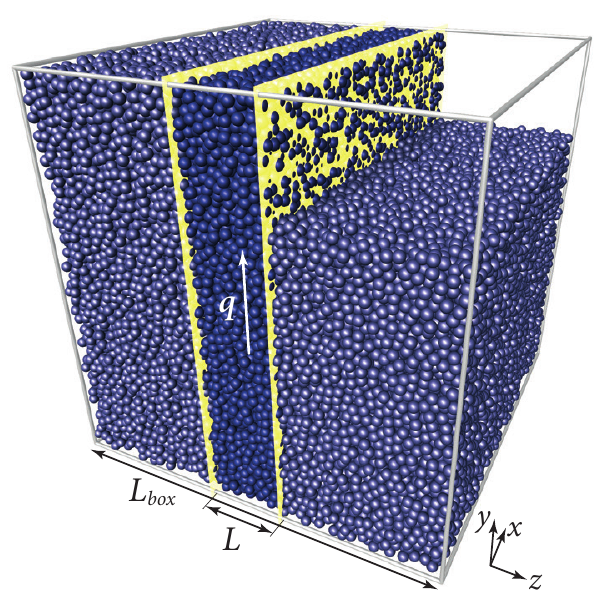}
  \caption{A liquid slab virtually cut out from a cube of liquid, creating open boundaries at the new surfaces (yellow), as opposed to the periodic boundary conditions at the faces of the cubic simulation domain (frame).
  The scattering wave vector $\vec q$ is parallel to the surfaces of the slab.
  Concerning the theoretical treatment, the thermodynamic limit $\Lbox \to \infty$ for fixed slab width $L$ is employed.
  The particles here are actually points, which have been assigned a non-zero size for illustrational purposes.
  Near the slab surfaces, the resulting, apparent cuts through particles underscore that the fluid structure is unchanged by the boundary. The position of the centre of a spherical particle determines whether it belongs to the slab or not.
  }
  \label{fig:sketch}
\end{figure}

Here, we discuss the static structure factor $S(|\vec q|; L)$ of a liquid slab delimited by two planar, open boundaries at a distance $L$. The sample can be thought of as \emph{virtually} cut out from a homogeneous fluid, so that no distortions occur near the surfaces (\cref{fig:sketch}).
Crucially, the wave vector $\vec k = (\vec q, 0)$ is chosen to be parallel to these surfaces in order to avoid interference with the finite extent in the perpendicular direction, which is chosen as the $z$-axis.
We shall derive a closed, exact expression for the static structure factor $S(|\vec q|; L)$ of this setup, which deviates significantly from the structure factor of the corresponding bulk phase.
In particular, the small-wavenumber limit $S(q \to 0; L)$ is shown to describe the (bulk) compressibility for wide slabs ($L\to \infty$) and particle number fluctuations of a small subsystem,
the latter having been discussed recently within micro-thermodynamics \cite{Schnell:CPL2011,Cortes-Huerto:JCP2016,Heidari:E2018}.

\section{Static structure factor of a liquid slab}
\label{sec:ssf_slab}

Consider a homogeneous and isotropic fluid of $\mathscr{N}$ point particles enclosed in a volume $V$, so that the number density is $\rho=\mathscr{N}/V$. The fluctuating three-dimensional positions of the particles are denoted by $\vec r_j$ ($j = 1, \dots, \mathscr{N}$).
Accordingly, the microscopic definition of the
static structure factor of this bulk phase reads \cite{Hansen:SimpleLiquids}
\begin{equation}
  \label{eq:ssf_def}
  S_b(|\vec k|) = \frac{1}{\mathscr{N}} \expect{|\hat \rho_{\vec k}|^2}
  \quad\text{with}\quad
  \hat\rho_{\vec k}:=\sum\nolimits_{j=1}^{\mathscr{N}} \e^{\i \vec k \cdot \vec r_j} \,,
\end{equation}
which is straightforward to evaluate in a simulation;
$\vec k$ is a three-dimensional wave vector and the quantities $\hat \rho_{\vec k}$ are Fourier modes of the fluctuating density field $\hat\rho(\vec r) = \sum_{j=1}^{\mathscr{N}} \delta(\vec r - \vec r_j)$.
For a finite simulation domain, the spatial homogeneity of the sample is effectively achieved by applying periodic boundary conditions along all Cartesian directions
and by restricting to wave vectors $\vec k$ of the reciprocal lattice of this periodically repeated domain (e.g., for a cubic box of edge length~$\Lbox$, each component of $\vec k$ is an integer multiple of $2\pi/\Lbox$).
At such $\vec k$, the density modes $\hat\rho_{\vec k}$ of an unbounded bulk sample are exactly resolved in the simulation, and the bulk structure factor $S_b(\vec k)$ obtained for periodic boundaries is not subject to finite-size corrections.

The slab structure factor $S(q; L)$ is defined as in \cref{eq:ssf_def} but with the sum restricted to particles in the slab
and the wave vector chosen as $\vec k = (\vec q, 0)$, where $\vec q$ is a two-dimensional vector.
Introducing the notation $\vec r_j = (\vec R_j, z_j)$ and denoting by $J$ the index set of particles for which $0 \leq z_j \leq L$,
it is given by
\begin{equation}
  S(|\vec q|;L) = \frac{1}{\expect{N}} \,\expect{\sum_{i,j \in J} \e^{\i\vec q \cdot(\vec R_i - \vec R_j)}} \,,
  \label{eq:ssf_slab_def}
\end{equation}
where $N$ is the number of particles in the slab for each sampled configuration;
clearly, $\expect{N} = \mathscr{N} L / \Lbox$.
We aim for deriving an expression for $S(q;L)$ in terms of the bulk structure factor, noting that the structure within the slab is entirely determined by the properties of the homogeneous liquid as the open boundaries do not distort the number density $\hat\rho(\vec r)$.
The idea is to express $S(q; L)$ in terms of general two-point density correlations, for a moment not exploiting the spatial homogeneity, and to relate the latter correlation function to $S_b(k)$, which closes the equations.

Within the theory of inhomogeneous fluids, one defines the density--density correlation function between two points $\vec r$ and $\vec r'$ in space as \cite{Evans:1979,Hansen:SimpleLiquids}
$
  G(\vec r, \vec r')=\expect{\hat\rho(\vec r) \, \hat\rho(\vec r')}
    - \expect{\hat\rho(\vec r)} \expect{\hat\rho(\vec r')}
$
with $\hat\rho(\vec r)$ as the microscopic number density and
with $\rho(\vec r) = \expect{\hat\rho(\vec r)}$ as the mean number density at point $\vec r$.
In planar geometry and in a statistical sense, the fluid is invariant under translations parallel to the slab surface. This entails
$\rho(\vec r) = \rho(z)$ and $G(\vec r, \vec r') = G(\vec R - \vec R', z, z')$,
and it suggests to use a lateral Fourier transform in the $xy$-plane,
where $\vec r = (\vec R, z)$.
For two-dimensional vectors $\vec q$ and $\Delta\vec R=\vec R - \vec R'$, one defines
\begin{align}
%   G(\vec q, z, z') \,\delta(\vec q-\vec q')
%     &= \int \diff^2 R \, \diff^2 R' \, \e^{-\i \vec q \cdot \vec R + \i \vec q' \cdot \vec R'} \, G(\vec R - \vec R', z, z') \\
  G(|\vec q|, z, z')
    &:= \int \diff^2 \Delta R \, \e^{-\i \vec q \cdot \Delta \vec R} \, G(\Delta \vec R, z, z') \,,
  \label{eq:Gq_def}
\intertext{which is equivalent to}
  G(|\vec q|, z, z')
    &= A^{-1} \Bigl[ \expect{\hat \rho_{\vec q}(z)^*\, \hat \rho_{\vec q}(z')}
      - \rho(z)\,\rho(z') \, \delta_{\vec q, \vec 0} \Bigr] \,
  \label{eq:Gq_def2}
\end{align}
in terms of the lateral density modes
\begin{align}
  \hat \rho_{\vec q}(z)
  &:= \int \! \diff^2 R \,\e^{\i \vec q \cdot \vec R} \, \hat\rho(\vec r = (\vec R, z)) \notag \\
  &= \sum\nolimits_{j=1}^{\mathscr{N}} \e^{\i \vec q \cdot \vec R_j} \delta(z - z_j)
\end{align}
with $A :=\int\!\diff^2 R = V / L$ as the area of one slab surface.
Substituting $\hat\rho_{\vec q}(z)$ into \cref{eq:Gq_def2}, we obtain the microscopic expression
$(\vec q\neq  0)$
\begin{equation}
  G(|\vec q|, z, z')
    = A^{-1} \, \expect{\sum_{i,j} \e^{-\i \vec q \cdot (\vec R_i - \vec R_j)} \delta(z-z_i) \delta(z'-z_j)} \,.
\end{equation}
The restriction of the particle sums in \cref{eq:ssf_slab_def} to the slab is then implemented by
\begin{equation}
  S(q;L) = \frac{1}{\rho L} \iint\limits_{\mathclap{0 \leq z, z' \leq L}} \diff z\,\diff z' \, G(q, z, z').
  \label{eq:gid_master_slab}
\end{equation}

The remaining task is to compute $G(q, z, z')$ for a homogeneous fluid.
To this end, it is favourable to introduce the pair distribution function $g(\vec r,\vec r')$,
which describes the so-called distinct part of the density correlation function; it fulfils \cite{Hansen:SimpleLiquids}
\begin{equation}
  G(\vec r, \vec r') = \rho(\vec r)\, \rho(\vec r') \, [g(\vec r, \vec r') - 1] + \rho(\vec r) \,\delta(\vec r - \vec r') \,.
  \label{eq:G_general}
\end{equation}
Due to lateral translational invariance one has $\rho(\vec r) = \rho = \mathit{const}$, and
$g(\vec r, \vec r')$ reduces to a function of the distance $|\vec r - \vec r'|$ only.
In this case, a three-dimensional Fourier transform uniquely links the pair distribution function and the bulk structure factor:
\begin{align}
  \rho\,g(\vec r, \vec r')
  &= \rho\,g(|\vec r - \vec r'|) \notag \\
  &= \int\!\!\frac{\diff^3 k}{(2\pi)^3} \,\e^{\i\vec k\cdot (\vec r - \vec r')} \, [S_b(|\vec k|) - 1] \,.
  \label{eq:ssf_sum_rule}
\end{align}
Replacing $g(\vec r, \vec r')$ in \cref{eq:G_general}
% \begin{equation}
%   G(\vec r, \vec r')
% %   = \rho^2 \, [g(\vec r - \vec r') - 1] + \rho \,\delta(\vec r - \vec r') \\
%   = \rho \! \int \!\! \frac{\diff^3 k}{(2\pi)^3} \,\e^{\i\vec k\cdot (\vec r-\vec r')} \, S(\vec k) - \rho^2 \,.
% \end{equation}
and writing for the wave vector $\vec k = (\vec q, k_z)$ invert the Fourier transform in the $xy$-plane [\cref{eq:Gq_def}] and yield the relation between $G(q, z, z')$ and $S_b(k)$:
\begin{multline}
  G(|\vec q|, z, z')
  = \rho \! \int \!\frac{\diff k_z}{2\pi} \,\e^{\i k_z (z - z')} \, S_b\Bigl(\!\sqrt{|\vec q|^2+k_z^2}\Bigr) \\
      - (2\pi \rho)^2\delta(\vec q) \,.
  \label{eq:Gq_bulk}
\end{multline}

Eventually, we combine \cref{eq:gid_master_slab,eq:Gq_bulk} and evaluate the integrals over $z$ and $z'$,
\begin{align}
  \iint\limits_{0 \leq z, z' \leq L} \hspace{-2.5ex} \diff z\,\diff z' \, \e^{\i k_z (z-z')}
  &= \left| \int_0^L \! \e^{\i k_z z} \diff z \right|^2 \notag \\
  &= \frac{2-2\cos(k_z L)}{k_z^2} \,,
  \label{eq:z-integral}
\end{align}
so that
\begin{multline}
  S(|\vec q|; L)
    = \frac{2}{\pi L} \! \int_0^\infty \!\diff k_z \,
      \frac{1-\cos(k_z L)}{k_z^2} \,
      S_b\Bigl(\!\sqrt{|\vec q|^2 + k_z^2}\Bigr) \\
      + \rho L (2\pi)^2 \delta(\vec q) \,.
  \label{eq:ssf_slab}
\end{multline}
This relation uses the bulk structure factor $S_b(k)$ as the only input for predicting the structure factor of a liquid slab with open boundaries.
Concerning the numerical evaluation of the integral over $k_z$, \cref{eq:ssf_slab} is recast into the form
\begin{multline}
  S(q > 0; L)
    = 1 + \frac{2}{\pi} \! \int_0^\infty \!
      \frac{1-\cos(x)}{x^2} \, \\
    \times \mleft[{\textstyle S_b\Bigl(\!\sqrt{q^2 + (x/L)^2}\Bigr)} - 1 \mright] \, \diff x,
  \label{eq:ssf_slab_numerics}
\end{multline}
employing the integral $\int_0^\infty x^{-2}[1-\cos(x)] \,\diff x = \pi / 2$.
For large $x$, the integrand decays rapidly, which facilitates the approximate truncation of the integration domain.
Furthermore, \cref{eq:ssf_slab_numerics} implies that the bulk structure factor is indeed recovered for an infinitely thick slab:
\begin{equation}
  S(q;L \to \infty) = S_b(q) \,,
  \label{eq:ssf_slab_limit}
\end{equation}
because the limit $L \to \infty$ can be interchanged with carrying out the integral.

We briefly consider the situation of a slab with periodic boundary conditions on the surfaces $z=0$ and $L$.
In this case,
the integral $(2\pi)^{-1} \int \!\diff k_z \cdots$ in \cref{eq:Gq_bulk} is replaced by the sum $L^{-1} \sum_n \cdots$ over discrete wavenumbers $k_z = 2\pi n/L$ for $n\in \mathbb{Z}$.
For such $k_z$, the integral $\int_0^L \e^{\i k_z z} \diff z$ vanishes except for $k_z = 0$.
Thus, the sum runs only over a single term, yielding $S_\text{per}(q;L) = S_b(q)$ instead of \cref{eq:ssf_slab}.
We conclude, that for periodic boundary conditions the slab structure factor is identical to the bulk one, irrespective of how small $L$ is.
As a consequence, for wide slabs ($L \to \infty$) the slab structure factors for open and periodic boundaries, respectively, approach each other [\cref{eq:ssf_slab_limit}], and we infer that the boundary condition becomes irrelevant in this limit.

\section{Simulation results}

\begin{figure*}
  \centering
  \includegraphics[scale=1]{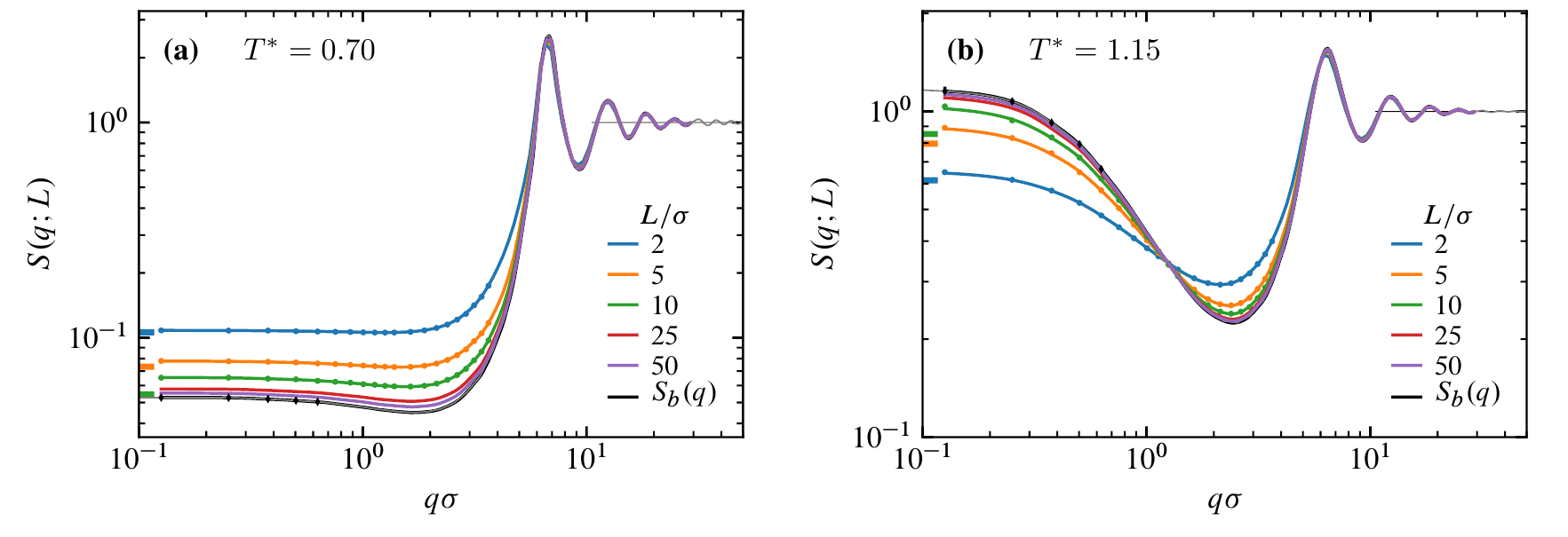}
  \caption{Structure factors $S(q;L)$ of liquid slabs of width $L$ with open boundaries in transverse direction, obtained from \cref{eq:ssf_slab} [see also \cref{eq:ssf_slab_asymptotics,eq:correction_integral}].
  These results are based on the bulk structure factor $S_b(q)$ of the Lennard-Jones liquid
  along the liquid--vapour coexistence line at temperatures $T^*=0.70$ [panel~(a)] and $T^*=1.15$ [panel (b)]; the corresponding densities are $\rho=0.824\sigma^{-3}$ and $0.540\sigma^{-3}$, respectively.
  The bulk structure factor $S_b(q)$ was obtained from simulations of a cubic system of volume $(50\sigma)^3$ with periodic boundaries along all Cartesian directions.
  The coloured bars at the vertical axis indicate simulation results for the Fano factor $F_N(L)$ of the particle number [\cref{eq:fano_def}], which are to be compared with the limiting values of $S(q \to 0; L)$.
  }
  \label{fig:ssf_bulk_slab}
\end{figure*}

As an example, we study the behaviour of $S(q; L)$ for Lennard-Jones (LJ) liquids at two temperatures along the liquid--vapour coexistence curve: $T^*=0.70 \approx T_t^*$ (close to, but slightly above the triple point) and $T^*=1.15$ ($\approx 94\%$ of the critical temperature $T_c^*$);
the corresponding densities of the liquid are $\rho=0.824\sigma^{-3}$ and $0.540\sigma^{-3}$, respectively \cite{Capillary:2015}.
In this study, the LJ pair potential was truncated at pair distances beyond $r_c = 3.5\sigma$, $T^* = \kB T/\epsilon$ denotes the reduced temperature, and $\epsilon$ and $\sigma$ are the LJ parameters for the interaction strength and range, respectively.
The bulk structure factors $S_b(k)$, serving as input to \cref{eq:ssf_slab}, were obtained according to \cref{eq:ssf_def,eq:ssf_slab_def} from massively parallel molecular dynamics simulations \cite{Glassy_GPU:2011, HALMDv1.0} (for details see Ref.~[\onlinecite{Capillary:2015}]).
We used a cubic simulation box of edge length $\Lbox=50\sigma$ with periodic boundaries on all faces;
at the higher density, it contained $\mathscr{N}=103{,}000$ particles.
Due to the finite extent of the box only wave vectors of the reciprocal lattice, $\vec k \in (2\pi/\Lbox)\mathbb{Z}^3$, are permissible.
Following the principle of data economy, memory transfer and hard disk access were greatly diminished by computing the structure factors online as the simulation was progressing and by storing the results as compressed, multi-dimensional data sets in the H5MD file format along with other simulation data \cite{H5MD:2014}.

For the numerical evaluation of \cref{eq:ssf_slab}, we used a parabolic spline interpolation of the simulated bulk structure factors $S_b(k)$ as function of $k^2$,
which was extended to large $k \gtrsim 30\sigma^{-1}$ by a poor man's hard-sphere expression for $S_b(k)$ in order to avoid overshoots of the spline at smaller~$k$; specifically, we utilised
$
  S_b(k) = 1 - 4 \pi a \rho k^{-3} [\sin(k \sigma) - k \sigma \cos(k\sigma)],
$
equivalent to a step function for $g(r)$, upon fitting the amplitude $a$ of the oscillations.
With this, the integral in \cref{eq:ssf_slab_numerics} was truncated at $k_\mathrm{max} = 50 /\sigma$ and evaluated by the routine \texttt{quad} from the \texttt{integrate} library of Scientific Python (SciPy), which wraps the Fortran library QUADPACK.

The obtained slab structure factors $S(q;L)$ are displayed in \cref{fig:ssf_bulk_slab}.
For small wavenumbers, $q\sigma \lesssim 4$, the figure exhibits a significant dependence of $S(q;L)$ on the slab width $L$.
For $L=5\sigma$, the value of $S(q\to 0;L)$ at the low temperature ($T^*=0.70$) is increased relative to its bulk value by about 50\%, whereas it is decreased by about 25\% at the higher temperature ($T^*=1.15$).
The residual small discrepancy between $S(q;L=L_\text{box})$ and the bulk structure factor $S_b(q)$ reflects the different boundary conditions and would disappear only in the limit $\Lbox \to \infty$.

\section{Analysis of the asymptotic behaviour}

Connecting to \cref{sec:ssf_slab}, it is straightforward to work out the asymptotic corrections to $S_b(q)$ due to a large, but finite slab ($L\to \infty$).
Rearranging \cref{eq:ssf_slab} similarly to \cref{eq:ssf_slab_numerics} yields
\begin{multline}
  S(q > 0; L) = S_b(q) +
    \frac{2}{\pi L} \! \int_0^\infty \! \diff k_z\,
      [1-\cos(k_z L)] \, \\
    \times \frac{\textstyle S_b\Bigl(\sqrt{q^2 + k_z^2}\Bigr) - S_b(q)}{k_z^2} \,.
  \label{eq:ssf_slab_numerics2}
\end{multline}
Note that the second factor of the integrand, $f(k_z) := \bigl[S_b\bigl(\sqrt{q^2 + k_z^2}\bigr) - S_b(q)\bigr]/k_z^2$, is bounded as $k_z \to 0$ due to
\begin{equation}
  S_b\Bigl(\sqrt{q^2 + k_z^2}\Bigr) = S_b(q) + \frac{k_z^2}{2q} S'(q) + O\bigl(k_z^4\bigr).
  \label{eq:ssf_small_kz}
\end{equation}
Furthermore, $f(k_z)$ is a function of $k_z^2$ by isotropy of the fluid and, away from a critical point, it is analytic in a disc around $k_z=0$, which implies an exponentially fast decay of the cosine transform \cite{Mimica:JMAA2016}:
\begin{equation}
  \int_0^\infty \! \cos(k_z L) f(k_z) \, \diff k_z
    = O\bigl(\e^{-L}\bigr) \quad \text{as \: $L \to \infty$} \,;
  \label{eq:ssf_residual}
\end{equation}
mathematically closely related situations are discussed in Refs.~[\onlinecite{Parry:2016,Parry:NP2019,Straube:2020}].
With that the expansion of $S(q;L)$ in terms of $L^{-1}$ follows as
\begin{equation}
  S(q>0;L) = S_b(q)  + 2 L^{-1} \mathcal{J}_0(q) + O\bigl(L^{-1}\e^{-L}\bigr) \,,
  \label{eq:ssf_slab_asymptotics}
\end{equation}
where we have introduced the integral
\begin{equation}
  \mathcal{J}_0(q) := \frac{1}{\pi} \int_0^\infty\! \diff k_z \,
    \frac{S_b\bigl(\sqrt{q^2 + k_z^2}\bigr) - S_b(q)}{k_z^2} \,,
  \label{eq:correction_integral}
\end{equation}
which depends on the bulk structure factor only.

\begin{figure}[b]
  \includegraphics[width=\figwidth]{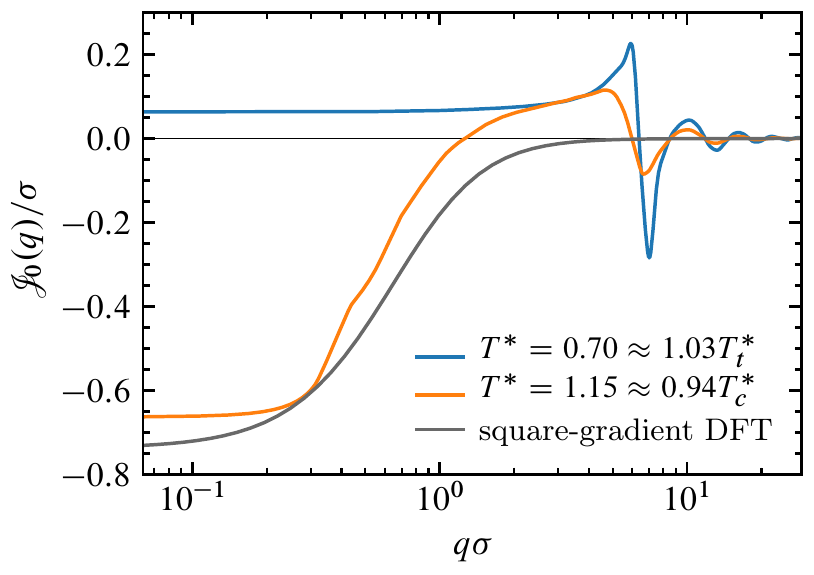}
  \caption{Leading finite-size correction $\mathcal{J}_0(q)$ of the slab structure factor [see \cref{eq:ssf_slab_asymptotics}] for the two LJ liquids analysed here.
  The coloured lines result from quadratures of \cref{eq:correction_integral} with the simulated bulk structure factors as input.
  The functional shape of $\mathcal{J}_0(q)$ as obtained within square-gradient DFT [\cref{eq:correction_integral_OZ}] is shown as a dark grey line with the parameters $S_0=1.18$ and $\xi=1.25\sigma$, set to their values for the LJ liquid at $T^*=1.15$.
  }
  \label{fig:I0}
\end{figure}

The behaviour of $\mathcal{J}_0(q)$ is illustrated for the simulated LJ liquids (see \cref{fig:I0}).
Close to the triple point, it is an essentially constant function of $q$ for not too large wavenumbers, i.e., $q\sigma \lesssim 3$.
For $T\approx 0.94 T_c$, however, $\mathcal{J}_0(q)$ increases monotonically from negative values and exhibits a change of sign.
In both cases, $\mathcal{J}_0(q)$ shows rapidly decaying oscillations at large $q$, picked up from the bulk structure factors.

The calculation of the integral in \cref{eq:correction_integral} is susceptible to details of the numerics and needs special care in two respects:
first, the slow decay $\propto k_z^{-2}$ of the integrand requires that a tail correction
$[1 - S_b(q)] / (\pi k_\text{max})$
is added to make up the integral for the truncation at $k_z = k_\text{max}$,
where we used that $S_b(\sqrt{q^2 + k_z^2})) \to 1$ for $k_z \geq k_\text{max} \gg \sigma^{-1}$.
Second, in order to avoid a spurious divergence of the integrand as $k_z \to 0$ it is essential to use a polynomial interpolation of $S_b(k)$ in terms of $k^2$ rather than $k$, which enforces the property $S_b'(0)=0$ demanded by the rotational invariance of the bulk phase.
Note that the small modulation in the data for $\mathcal{J}_0(q)$ around $4\lesssim q\sigma \lesssim 7$ (\cref{fig:I0}) is likely to be a numerical artifact, which we attribute to the aforementioned subtleties.

Some theoretical insight is gained by assuming that $S_b(k)$ is of the Ornstein--Zernike (OZ) form,
$
  S_b(k) = S_0 / \bigl[1 + (k\xi)^2\bigr]\,,
$
which follows from square-gradient density-functional theory (DFT) and which is a reliable description of $S_b(k)$ for $k \xi \ll 1$ close to the critical point; $\xi$ denotes the OZ correlation length,
characterising the decay length of the two-point correlation functions.
For this ansatz, the integral in \cref{eq:correction_integral} can be carried out and yields
\begin{equation}
  \mathcal{J}_0(q) = -\frac{(\xi/2) S_0}{\bigl(1 + \xi^2 q^2\bigr)^{3/2}}\,.
  \label{eq:correction_integral_OZ}
\end{equation}
For the residual integral [\cref{eq:ssf_residual}] we obtain
\begin{equation}
  \int_0^\infty \!\diff k_z \, \cos(k_z L) f(k_z) = \pi \mathcal{J}_0(q) \, \e^{-L\sqrt{q^2+\xi^{-2}}} \,.
\end{equation}
This indeed renders an exponentially fast decay for slab widths $L$ which are larger than either the correlation length~$\xi$ or the wavelength $2\pi/q$.
Within this simple model, the finite-size correction $\propto \mathcal{J}_0(q)$ of the slab structure factor is strictly negative and monotonically increasing from $\mathcal{J}_0(q \to 0) = -(\xi/2) \, S_0$ towards zero at large~$q$.
In spite of its simplicity, \cref{eq:correction_integral_OZ} can be considered as a useful approximation for actual fluids close to criticality (\cref{fig:I0}).

Close to the triple point, $T_t^* \approx 0.70$ [\cref{fig:ssf_bulk_slab}(a)], both the compressibility of the liquid and the correlation length are small ($S_0 \ll 1$ and $\xi \ll \sigma$, respectively),
which leads to $\mathcal{J}_0(q \to 0) > 0$ (see \cref{fig:I0}).
The latter can be understood by noticing that here $S_b(0)$ acts as an approximate lower bound on $S_b(k)$,
which suggests that the integrand in \cref{eq:correction_integral} is dominated by positive values for $q \to 0$.
Interestingly, there is a distinguished temperature (along the liquid--vapour coexistence curve) at which the small-$q$ correction vanishes, i.e., $\mathcal{J}_0(q \to 0) = 0$.

\section{Compressibility and fluctuations of the particle number}

The small-wavenumber value of the structure factor is a measure of the isothermal compressibility $\chi_T^\infty$ of the fluid \cite{Hansen:SimpleLiquids}:
\begin{equation}
  S_b(k \to 0) = \rho \kB T \chi_T^\infty \,,
\end{equation}
where the superscript $\infty$ indicates a macroscopically large sample.
This relation is usually derived in the grand canonical ensemble by starting from the thermodynamic definition of $\chi_T^\infty$ and showing that the r.h.s.\ equals the Fano factor (sometimes also referred to as the index of dispersion) of the fluctuating particle number:
\begin{equation}
  \rho \kB T \chi_T^\infty = \frac{\Var[\mathscr{N}]}{\expect{\mathscr{N}}} =: F_{\mathscr{N}}^\infty \,,
\end{equation}
where $\Var[\mathscr{N}] = \expect{\mathscr{N}^2} - \expect{\mathscr{N}}^2$ denotes the variance of $\mathscr{N}$.
A Fano factor $F_{\mathscr{N}} \neq 1$ quantifies how the distribution of $\mathscr{N}$ deviates from a Poisson distribution, here corresponding to the case of an ideal gas.

Moreover, one has that $S_b(k \to 0) = F_{\mathscr{N}}^\infty$ irrespective of the statistical ensemble,
with an analogous sum rule applying for the slab structure factor $S(q;L)$.
The standard proof \cite{Hansen:SimpleLiquids} is based on integration of \cref{eq:ssf_sum_rule} over $\vec r, \vec r'$ and counting particles. Here, we present this proof for $S(q;L)$ in a condensed form.
From \cref{eq:ssf_slab_def}, one obtains
\begin{multline}
  \int\!\!\frac{\diff^2 q}{(2\pi)^2} \,\e^{\i\vec q\cdot \vec R} \, S(|\vec q|; L)
    = \frac{1}{\expect{N}} \, \expect{\sum_{i \neq j} \delta(\vec R_i - \vec R_j - \vec R)} \\
      + \delta(\vec R) \,,
\end{multline}
where the term $\delta(\vec R)$ results from the self part ($i = j$).
Including the latter in the l.h.s.\ and integrating over $\vec R$, we find
\begin{equation}
  \int \! \diff^2 R\int\!\!\frac{\diff^2 q}{(2\pi)^2} \,\e^{\i\vec q\cdot \vec R} \, [S(|\vec q|; L) - 1]
    = \frac{\expect{N (N - 1)}}{\expect{N}} \,.
  \label{eq:chi_derivation1}
\end{equation}
On the other hand, carrying out the $\vec R$-integral first, the l.h.s.\ turns into
\begin{equation}
  \int\!\diff^2 q\,\delta(\vec q) \, [S(|\vec q|; L) - 1]
    = S(q \to 0; L) - 1 + \rho L A \,,
  \label{eq:chi_derivation2}
\end{equation}
where the first term refers to the continuous extension of $S(q;L)$ to $q=0$.
The singular peak at $\vec q=0$ [see \cref{eq:ssf_slab}] must be handled separately and generates the last term, i.e., $\rho L A = \expect{N}$, for a large, but finite slab area~$A$.
Equating the r.h.s. of both \cref{eq:chi_derivation1} and \cref{eq:chi_derivation2} proves the sum rule for the slab:
\begin{equation}
  S(q \to 0; L) = \frac{\Var[N]}{\expect{N}} =: F_N(L) \,.
  \label{eq:fano_def}
\end{equation}

\begin{figure}
 \includegraphics[width=\figwidth]{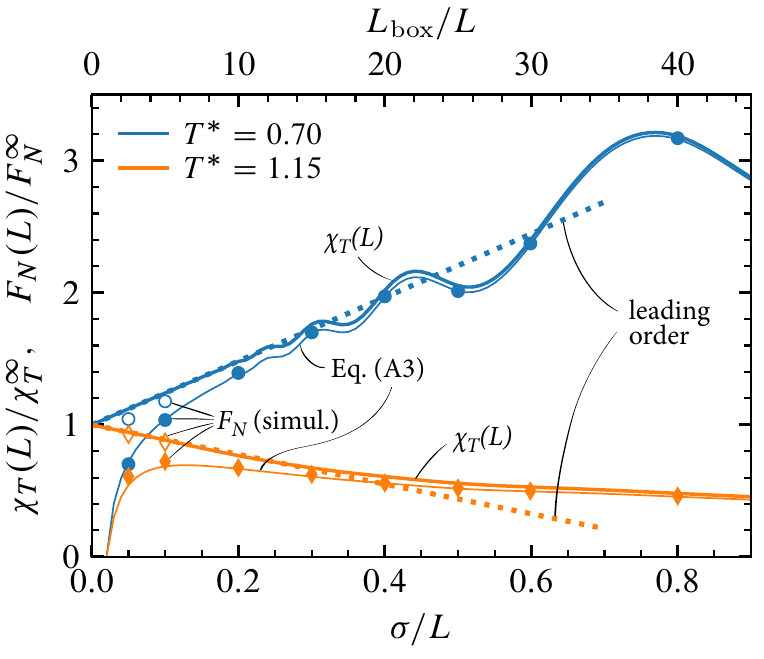}
 \caption{
  Compressibility $\chi_T(L)$ of the slab (thick lines) and simulation data for the Fano factor $F_N(L)$ of the fluctuating particle number (symbols) as functions of the inverse slab width~$1/L$;
  the results are shown for the two thermodynamic state points (blue and red) studied before (see \cref{fig:ssf_bulk_slab}).
  The slab compressibility is defined via the small-$q$ limit of the slab structure factor [\cref{eq:chi_L_def}] and thick solid lines are predictions from the numerical calculation of \cref{eq:chi_L} with the bulk structure factors as input;
  dotted straight lines indicate their large-$L$ asymptotes [\cref{eq:chi_L_asymptotics}].
  The Fano factors $F_N(L)$ were obtained directly from the statistics of simulated particle counts in the slab for a cubic simulation domain of fixed size $\Lbox=50\sigma$ (full symbols, filled circles and diamonds); the error bars are much smaller than the symbol size.
  Open symbols (circles and diamonds) show the results for cuboid boxes with an eightfold increased length perpendicular to the slab, $\Lbox^{(z)} = 400\sigma$.
  This increase of $\Lbox$ facilitates that the full symbols are shifted upwards and thus turn into open symbols, which follow closely the expected large-$L$ behaviour, i.e., approaching~1.
  Thin lines refer to the prediction of \cref{eq:FN_Lbox} for $F_N(L;\Lbox=50\sigma)$, which accounts also for the finite size of the simulation box. In particular, \cref{eq:FN_Lbox} correctly yields $F_N=0$ for $L =\Lbox$, because in that limit the number of particles does not fluctuate.
  All data are normalised by their bulk values in the grand canonical ensemble, i.e., $\chi_T^\infty$ and $F_N^\infty=\rho \kB T \chi_T^\infty$, respectively,
  so that $\chi_T(L) / \chi_T^\infty = F_N(L) / F_N^\infty$.
  }
 \label{fig:compressibility}
\end{figure}

A grand canonical ensemble is realised if the slab occupies a volume which is small compared to the remaining part of the simulation box, so that the latter can act as a reservoir for the open subsystem.
This suggests to introduce the isothermal compressibility $\chi_T(L)$ of the slab by virtue of
\begin{equation}
  S(q \to 0; L) =: \rho \kB T \chi_T(L) \,,
  \label{eq:chi_L_def}
\end{equation}
which in general differs from the bulk compressibility $\chi_T^\infty$;
from \cref{eq:fano_def} it follows $F_N(L) = \rho \kB T \chi_T(L)$.
The discrepancy between $\chi_T^\infty$ and $\chi_T(L$) is evident from our analytical results in \cref{eq:ssf_slab,eq:ssf_slab_asymptotics}, which imply
\begin{align}
  \rho \kB T \chi_T(L)
    &= 1 + \frac{2}{\pi L} \! \int_0^\infty \!\diff k_z \,
      \frac{1-\cos(k_z L)}{k_z^2} \, [S_b(k_z) - 1]
  \label{eq:chi_L}
\intertext{so that for $L\to\infty$ one has [compare with \cref{eq:ssf_slab_asymptotics}]}
  \chi_T(L) &\simeq \chi_T^{\infty} + \frac{2 \mathcal{J}_0(q \to 0)}{\rho \kB T L} + O\bigl(L^{-1}\e^{-L/\xi}\bigr) \,.
  \label{eq:chi_L_asymptotics}
\end{align}
\Cref{fig:compressibility} depicts the non-monotonic behaviour of $\chi_T(L)$ as predicted by the numerical calculation of \cref{eq:chi_L}.
The finite-size correction to $\chi_T^\infty$ scales with the linear dimension of the slab as $L^{-1}$, in line with a heuristic argument \cite{Rovere:1990} and numerical evidences \cite{Schnell:CPL2011,Cortes-Huerto:JCP2016,Heidari:E2018,Rovere:1993} for cubic subsystems.

Within the simulations, we have evaluated the Fano factor $F_N(L)$ from counting particles in the slab for different widths~$L$ (\cref{fig:compressibility}); this operation is well defined at the boundaries because LJ particles are point-like objects.
These simulation estimates are systematically below the theoretical value for $S(q\to 0; L) = F_N(L)$, but approach the latter for small $L$ and reproduce even the pronounced oscillations for narrow slabs and $T^*=0.70$.
We emphasise that the derivation of the slab structure factor in \cref{sec:ssf_slab} assumes that the slab is cut out from a macroscopically large, truly homogeneous liquid.
In particular, it does not include the periodicity with $\Lbox$ which a finite simulation box with periodic boundary conditions imposes on the density correlations.
Clearly, the (pathological) limit $F_N(L \to \Lbox) = 0$ is not contained in our theoretical results, which explains the increasing gap in \cref{fig:compressibility} between the predicted (solid lines) and observed (full symbols) particle statistics as $L$ approaches $\Lbox$.
The gap is almost closed by using a simulation box which is eightfold enlarged along the $z$-axis, i.e., $\Lbox^{(z)} = 400\sigma$ (open symbols).
This limitation of the theory can be exploited to infer the minimal ratio $\Lbox/L$ for which theory and simulations (see above) still nearly coincide, which may serve as a criterion how closely the simulated open subsystem models a grand canonical ensemble.
The data in \cref{fig:compressibility} suggest $\Lbox/L \gtrsim 20$ or $L / \Lbox \lesssim 0.05$ for the two fluids studied here.

The corrections due to a finite, periodic simulation box can be accounted for by repeating the derivation of \cref{eq:chi_L} as explained in \cref{sec:finite-size}, which leads to \cref{eq:FN_Lbox}.
The latter provides an accurate description of the Fano factor data for $\Lbox=50\sigma$ (\cref{fig:compressibility}, thin lines).
In a nutshell, the modifications to obtain \cref{eq:FN_Lbox} amount to, first, replacing the integral over $k_z$ by a sum over a discrete set of wavenumbers and, second, discarding the mode $k_z=0$ in order to implement the conservation of the total particle number in the system.
For sufficiently large boxes, i.e., $\Lbox \gg \sigma$, the sum in \cref{eq:FN_Lbox} can be approximated by reverting it to an integral over $k_z$ again, where attention must be paid to the missing term for $k_z=0$.
Comparison with \cref{eq:chi_L} yields the simple, approximate formula
\begin{equation}
  F_N(L, \Lbox) \approx F_N(L) - \frac{L}{\Lbox} F_N^\infty \,.
  \label{eq:FN_approx}
\end{equation}
At both temperatures studied, the evaluation of this expression for $\Lbox=50\sigma$ is visually indistinguishable from the exact result in \cref{eq:FN_Lbox} shown in \cref{fig:compressibility}
as thin lines.

% comment on correlation length, quote value ($\xi \approx 1.4\sigma$ for $T^*=1.15$), here $L$ is sufficiently large (?!)

\section{Summary and Conclusions}

Motivated by the availability of grazing-incidence X-ray scattering at planar interfaces of coexisting liquid and vapour phases, we studied the implications of open boundary conditions for a slab-shaped sample of an otherwise homogeneous liquid.
As the main observable, we introduced the static structure factor $S(q;L)$ of a liquid slab, describing lateral density fluctuations, i.e., with wave vectors lying parallel to the slab surfaces [\cref{eq:ssf_slab}].
The first result is an exact integral expression for $S(q;L)$, which requires only the bulk structure factor $S_b(k)$ of the homogeneous fluid as input [\cref{eq:ssf_slab}].
The expression was exemplified for and corroborated by simulation data of truncated LJ liquids at two thermodynamic state points, one close to the triple point and one near the liquid--vapour critical point (\cref{fig:ssf_bulk_slab}).
The asymptotic analysis of $S(q;L)$ for large slab widths $L$ shows that the difference between slab and bulk structure factors is accurately captured by the expression $2L^{-1} \mathcal{J}(q)$ [see \cref{eq:ssf_slab_asymptotics,eq:correction_integral}] and vanishes algebraically as $L$ increases.
The residual approximation error decays exponentially for $L$ larger than the correlation length $\xi$ of the fluid, which is a consequence of $S_b(k)$ being an analytic function in~$k^2$.
The finite-size correction integral $\mathcal{J}(q)$ does not depend on geometric parameters and is determined by $S_b(k)$ alone, which is routinely accessible to both simulations and experiments.
$\mathcal{J}(q)$ is an increasing function for not too large wavenumbers, but it can be of either sign and also exhibit a zero crossing (\cref{fig:I0}).
We emphasise that for \emph{periodic} boundary conditions on the slab surfaces, the slab structure factor
is not subject to finite-size corrections; rather, it identically resembles the bulk structure factor: $S_\text{per}(q;L) = S_b(q)$.

An important observation is that the slab structure factor is a non-additive function of~$L$, i.e.,
\begin{equation}
  L_1 S(q;L_1) + L_2 S(q;L_2) \neq (L_1 + L_2) \, S(q;L_1 + L_2)  \,,
\end{equation}
at variance with the periodic case $S_\text{per}(q;L)$, which is actually independent of~$L$.
Thus, $S(q;L)$ contains transverse correlations between particles at different $z$ positions
(i.e., between the two volumes of thickness $L_1$ and $L_2$), which are entirely discarded within the
approximation $S(q;L) \approx S_b(q)$.
The presence or absence of these correlations significantly affects results for the wavenumber-dependent surface tension, in particular at small wavenumbers and low temperatures, with the potential to flip the sign of the so-called bending coefficient \cite{Capillary:2015}.
Similarly, the quantity $\mathcal{J}_0(q)$ is crucial for the interpretation of GIXRD scattering data, aiming at an unambiguous separation of interfacial correlations from the background of homogeneous bulk phases \cite{GID:2020}.

Concerning the theory of inhomogeneous fluids within planar geometry, the study of the \emph{local} (or transverse) structure factor \cite{Tarazona:1982}
$S_\text{loc}(q,z) = \int \diff z' \, G(q, z,z')$
has recently proven to be very fruitful \cite{Parry:2014,Parry:2016,Evans:JPCM2015,Parry:NP2019}.
If the integral over $z'$ extents over the whole space (i.e., not only the slab), inspection of our derivation of $S(q;L)$ in \cref{sec:ssf_slab} shows that $S_\text{loc}(q,z) = S_b(q)$ for all $z$ within the slab, irrespective of whether open or periodic boundary conditions are imposed on the surfaces.
[In \cref{eq:z-integral}, the integral over $z'$ would yield $\delta(k_z)$.]
However, if the domain of $z'$ is restricted to the slab as well, the resulting expression for $S_\text{loc}(q,z;L)$ is a non-trivial integral of $S_b(q)$ with a finite-size correction scaling as $L^{-1}$, similarly to \cref{eq:ssf_slab,eq:ssf_slab_asymptotics}.

Eventually, through the open boundaries the liquid slab is coupled to the exterior fluid, which acts as a reservoir of particles.
We have shown that the Fano factor $F_N(L) = \Var[N] / \expect{N}$ of the fluctuating particle number in this open subsystem
is equal to the small-wavenumber limit $S(q \to 0;L)$ of the slab structure factor.
This sum rule provides a microscopic expression for $F_N(L)$, which is exact for all widths $L$, provided that the reservoir is sufficiently large, i.e., $\Lbox \gg L$.
In this case, the subsystem is expected to realise a grand canonical ensemble, which suggests the formal definition of an isothermal compressibility $\chi_T(L)$ of the slab in terms of $S(q \to 0;L)$.
Our results for $S(q;L)$ carry over to $\chi_T(L)$, yielding again an explicit formula [\cref{eq:chi_L}] and the asymptotic structure for large~$L$ [\cref{eq:chi_L_asymptotics}].
In particular, $\chi_L(T)$ deviates from the bulk compressibility, either increasing or decreasing, depending on the sign of $\mathcal{J}_0(q \to 0)$, and with the finite-size correction scaling as $L^{-1}$.
Our analytical findings for $\chi_L(T)$ and thus $F_N(L)$ are corroborated by data from large-scale simulations with $\Lbox \gtrsim 20 L$.
We anticipate that the setup of a liquid slab with open boundaries can serve as a meaningful test-bed for the thermodynamics of small, open systems and the simulations thereof \cite{DelleSite:2019,DelleSite:JSM2017,Schnell:CPL2011,Cortes-Huerto:JCP2016,Heidari:E2018}.
The presented approach suggests that for cubic subvolumes an analogous route could be followed to theoretically analyse the particle statistics as well as further observables.

\appendix

\section{Finite-size corrections due to periodic simulation boxes}
\label{sec:finite-size}

In actual computer simulations, the liquid slab is taken as a subvolume of a periodically repeated, finite chunk of fluid (see \cref{fig:sketch}), not of an infinitely extended fluid as assumed in the derivation carried out in the main text.
Therefore, the geometry is controlled by two length scales, the slab width $L$ and the edge length $\Lbox$ of the simulation box.
The periodicity of the fluid enters our derivation at the level of the pair correlation function.
Accordingly, one has to replace its expression for a homogeneous fluid [\cref{eq:ssf_sum_rule,eq:Gq_bulk}] by one which is periodic along the $z$-axis,
i.e., $G(q,z,z') = G(q,z + \Lbox,z')$ in addition to translational invariance, i.e., $G(q,z,z') = G(q,z+a,z'+a)$ for any shift $a$.
The former is achieved by replacing the Fourier integral $\int \cdots \diff k_z/2\pi$ by the discrete sum $\Lbox^{-1} \sum_{k_z} \cdots$ over wave vectors $k_z \in (2\pi/\Lbox) \mathbb{Z}$.
In particular, changing \cref{eq:Gq_bulk} to
\begin{equation}
  G(q > 0, z, z')
  = \frac{\rho}{\Lbox} \sum_{k_z}\e^{\i k_z (z - z')} \, S_b\Bigl(\!\sqrt{q^2+k_z^2}\Bigr)
  \label{eq:Gq_bulk_Lbox}
\end{equation}
propagates through the entire derivation.

In a (micro-)canonical simulation, the particle number is conserved and thus of zero variance, which requires
\begin{equation}
  \iint\limits_{0 \leq z, z' \leq \Lbox} \hspace{-2.5ex} G(q \to 0, z, z') \, \diff z\,\diff z' = 0.
\end{equation}
Combining this with the above form of $G(q,z,z')$ [\cref{eq:Gq_bulk_Lbox}], the integral renders zero for all discrete wavenumbers $k_z \in (2 \pi / \Lbox) \mathbb{Z}$, except for the $k_z=0$ mode, which in the limit $q \to 0$ must be excluded from the sum in \cref{eq:Gq_bulk_Lbox} in order to implement the constraint.

Adapting \cref{eq:ssf_slab,eq:chi_L} accordingly, we obtain for the Fano factor of the slab
% \begin{multline}
%   S(q > 0; L, \Lbox)
%     = \frac{L}{\Lbox} \, S_b(q) \\
%       + \frac{4L}{\Lbox} \sum_{k_z > 0} \,
%       \frac{1-\cos(k_z L)}{(k_z L)^2} \,
%       S_b\Bigl(\!\sqrt{q^2 + k_z^2}\Bigr) \,,
%   \label{eq:ssf_slab_Lbox}
% \end{multline}
\begin{equation}
  F_N(L, \Lbox) = \frac{4L}{\Lbox} \sum_{k_z > 0} \, \frac{1-\cos(k_z L)}{(k_z L)^2} \, S_b(k_z) \,.
  \label{eq:FN_Lbox}
\end{equation}
For $L=\Lbox$, every term of the sum vanishes due to $\cos(k_z \Lbox)=1$, so that $F_N(\Lbox, \Lbox) = 0$ as requested.
For the purpose of numerical evaluation, we rearrange \cref{eq:FN_Lbox} into the more rapidly converging form
\begin{multline}
  F_N (L,\Lbox) = 1 - \frac{L}{\Lbox}  \\ 
    + \frac{4L}{\Lbox} \sum_{k_z > 0} \, \frac{1-\cos(k_z L)}{(k_z L)^2} \, [S_b(k_z) - 1] \,.
\end{multline}
The equivalence of the last two equations follows from the identity
\begin{equation}
 \frac{1}{\Lbox}\sum_{k_z} \frac{1-\cos(k_z L)}{k_z^2 L^2} 
    = \int_{-\infty}^\infty \frac{\diff k_z}{2\pi} \frac{1-\cos(k_z L)}{k_z^2 L^2} \,,
\end{equation}
which can be obtained from the Euler--MacLaurin summation formula by evaluating the remainder term [see Eq. (23.1.32) in Ref.~\citenum{Abramowitz:MathFunc}].
% I made precise numerical estimates of the remainder integral using Mathematica, with $R_p \approx O(10^{-p})$ up to $p=20$.
Singling out the term for $k_z=0$ on the l.h.s., performing the integral on the r.h.s., and multiplying by $2L$, one finds
\begin{equation}
 \frac{4 L}{\Lbox}\sum_{k_z > 0} \frac{1-\cos(k_z L)}{k_z^2 L^2} = 1 - \frac{L}{\Lbox} \,.
\end{equation}

\begin{acknowledgments}
This research has been supported by Deutsche Forschungsgemeinschaft (DFG) through grant SFB~1114, project no.\ 235221301, sub-project C01.
The data that support the findings of this study are available from the corresponding author upon reasonable request.
%
% discussion/correspondence with Robert Evans?
\end{acknowledgments}

\bibliography{capillary}

\end{document}